\documentclass[prl,aps,10pt,twocolumn]{revtex4-1}
\usepackage{epsfig,graphicx,amsmath}
\usepackage{amsfonts}
\usepackage{amsmath}
\usepackage{enumerate}		
\usepackage{amsfonts}
\usepackage{amssymb}
\usepackage{amsthm}

\usepackage{afterpage}
\usepackage{color}

\usepackage[caption=false]{subfig}
\newcommand{\be}{\begin{equation}}
\newcommand{\ee}{\end{equation}}
\newcommand{\bea}{\begin{eqnarray}}
\newcommand{\eea}{\end{eqnarray}}

\newcommand{\ba}{\bea \begin{array}}
\newcommand{\ea}{\end{array} \eea}

\newcommand{\bc}{\begin{center}}
\newcommand{\ec}{\end{center}}

\newcommand{\beq}{\begin{eqnarray}}
\newcommand{\eeq}{\end{eqnarray}}
\newcommand{\beqq}{\begin{eqnarray*}}
\newcommand{\eeqq}{\end{eqnarray*}}

\newcommand{\Om}{\mbox{\boldmath$\omega$}}

\newcommand{\U}{\mbox{\boldmath$U$}}
\newcommand{\V}{\mbox{\boldmath$V$}}
\newcommand{\R}{\mbox{\boldmath$R$}}
\newcommand{\E}{\mbox{\boldmath$\eta$}}
\newcommand{\Id}{\mbox{\boldmath$I_d$}}
\newcommand{\M}{\mbox{\boldmath$M$}}
\newcommand{\1}{\mbox{\boldmath$1$}}

\begin{document} 
	\title{Randomly cross-linked polymer models}
	\author{O. Shukron and D. Holcman}
	\affiliation{Institute of Biology, Ecole Normale Sup\'erieure, 46 rue d'Ulm 75005 Paris, France.}
	
\begin{abstract}
Polymer models are used to describe chromatin, which can be folded at different spatial scales by binding molecules.  By folding, chromatin generates loops of various sizes. We present here a randomly cross-linked (RCL) polymer model, where monomer pairs are connected randomly. We obtain asymptotic formulas for the steady-state variance, encounter probability, the radius of gyration, instantaneous displacement and the mean first encounter time between any two monomers. The analytical results are confirmed by Brownian simulations. Finally, the present results can be used to extract the minimum number of cross-links in a chromatin region from {conformation capture} data.
\end{abstract}

\maketitle
\begin{figure}[htbp!]
		\centering
		\includegraphics[width=1\linewidth]{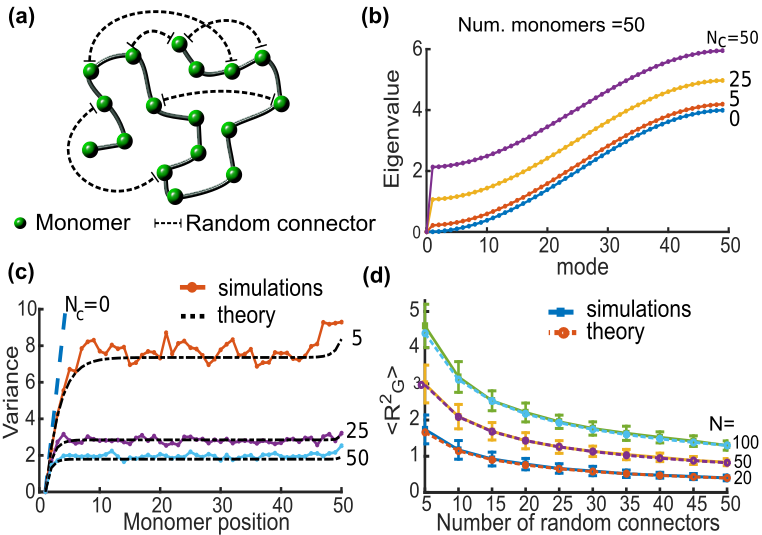}
		\caption{Steady-state properties of a Randomly-Cross-Linked (RCL) polymer. \textbf{(a)} representation of a RCL polymer composed of a linear backbone of $N$ monomers (spheres), and $N_c$ random connectors (dashed) between non-nearest neighboring monomers pairs. \textbf{(b)} Eigenvalues of the RCL polymer (Eq. \ref{eq:eigenValuesRCLPolymer}) with $N=50$ monomers, and $N_c=5$ (blue), 25 (red), and 50 (yellow) connectors. \textbf{(c)} Variance of the monomers' distance: analytical (dashed, Eq.\ref{eq:varianceExactForm}) versus simulated (Eq.\ref{eq:systemODENormalCoordinates}) between monomer 1 and monomers 2-50 of the RCL polymer, 500 realizations with $N=50, b=\sqrt{d},D=1,\Delta t=0.01s$, for $N_c=5$ (blue), 25 (yellow) and 50 (green) added random connectors. \textbf{(d)} The mean square radius of gyration, $\langle R_G^2(\xi)\rangle$, obtained from simulations of RCL polymers with $N=20$ (blue), 50 (yellow), and 100 (green) monomers: analytical (dashed, Eq. \ref{eq:MSRG}), where $N_c\in [5,50]$, compared to stochastic simulations  of system \ref{eq:systemODENormalCoordinates} (continuous).}
		\label{fig:Figure_1}
	\end{figure}
\begin{figure}[http!]
		\centering
		\includegraphics[width=1\linewidth]{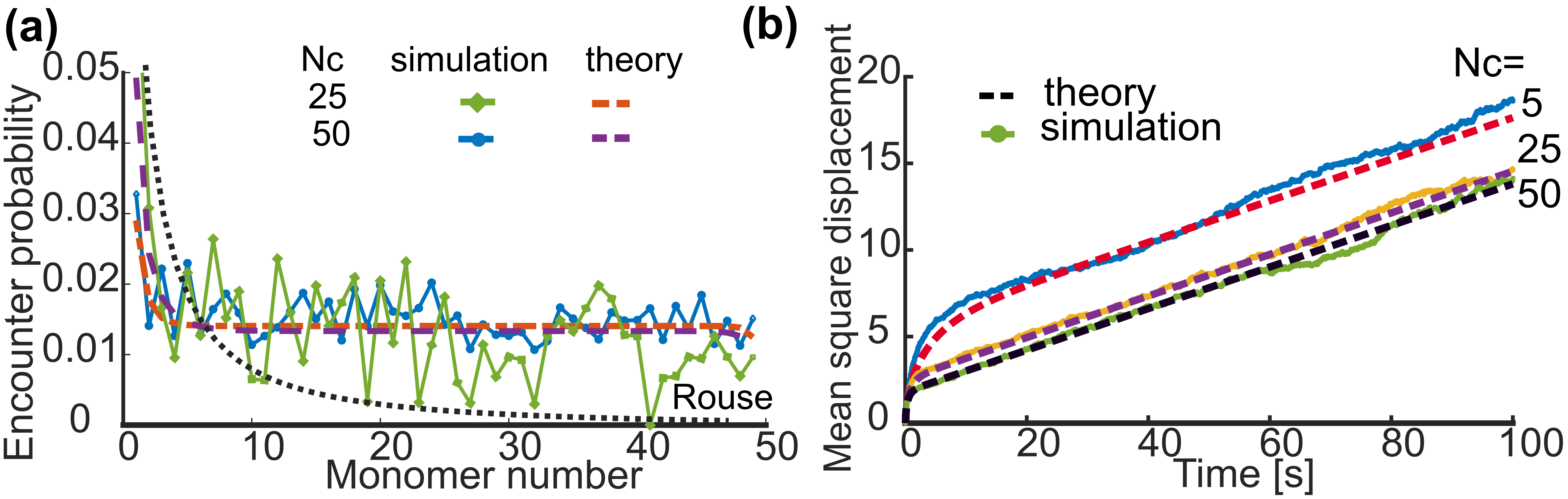}
		\caption{Dynamic properties of the RCL polymer \textbf{(a)} Simulation of encounter probability between monomers 1 and 2-50 of the RCL polymer.  with $N=50$ monomers, where we added $N_c=25$ (green diamonds) and 50 (blue circles) connectors and compared with analytical formula Eq. \ref{eq:encounterProbInRandomLoopModel} (dashed curves) for 500 realizations with $D=1, b=\sqrt{3}, \epsilon = b/10, \Delta t=0.01s$. We show the encounter probability of the Rouse polymer ($N_c=0$ connectors, dotted black), which cannot account for long-range connectivity. \textbf{(b)} Simulation of the mean square displacement for a RCL polymer of $N= 50$ monomer for $N_c=5$ (blue), 25 (yellow), and  50 (green) added connectors, shows excellent agreement with analytical formulas (Eq. \ref{eq:msdRCL},  dashed).}
\label{fig:Figure_2}
\end{figure}

DNA in the nucleus is constantly remodeled by regulatory factors and compacted genomic regions form transient and stable loops \cite{Nora2012,Tark-Dame2014}.  Looping is thus a key event in chromatin regulation: it is rare for a single polymer but frequent in a population of hierarchy folded genome. Genome organization is now probed by chromatin Conformation Capture (CC) techniques \cite{Dekker2002,Simonis2006,Lieberman2009}, which simultaneously give access to looping events in an ensemble of millions of chromatin segments. This experimental approach provides contact frequency matrices at various scale from few kilo- to Mega-base-pairs. Analysis of these matrices remains difficult, but revealed that mammalian genomes contain "blocks" of up to few Mbp in size, called Topologically Associating Domains (TADs) \cite{Nora2012,Dixon2012}. The role of TADs and organization remains unclear, although they are involved in gene regulation \cite{Nora2012,Simonis2006} and replication. TADs appear by averaging encounters over an ensemble of millions of samples \cite{Lieberman2009} and represents steady-state looping frequencies, but does not contain neither directly information about the size of the folded genomic section nor any transient genomic encounter times. \\
To reconstruct chromatin at a given scale and explore its transient properties, polymer models are used as a coarse-grained representation. The  Rouse model \cite{Doi1986}, characterized by nearest neighbors interactions, predicts an encounter probability (EP) that decays with $|m-n|^{-3/2}$  between monomer $m$ and $n$, but cannot account for long-range interactions, observed inside TADs of the CC data \cite{Shukron2017,Nora2012}. Other polymer models include attractive and repulsive forces between monomers \cite{Sokolov2003,Bohn2007,Bohn2010,Heermann2011,Amitai2013,Langowski2007} to account for long-range interactions and have been used to probe the heterogeneous steady-state organization  of the chromatin, characterized by loops \cite{Jost2014}. \\
We study here a randomly cross-linked (RCL) polymer model used in \cite{Shukron2017} to describe the ensemble of chromatin conformation, where random cross-links are formed by binding molecules \cite{Nora2012}. The space of RCL polymer configurations was so far explored only numerically \cite{Heermann2011,Jespersen2000}, but no analytical formulas have been derived for studying the steady-state or transient properties and thus explore the large parameter space. We derive here novel analytical formulas for the encounter probability (EP), variance, and the radius of gyration of the RCL polymer, that we use to study the polymer dynamics. The present model can be used to determine from CC empirical EP , the minimal number of cross-links, inaccessible from CC experiments. We further study the mean first encounter time between any two monomers, which plays a key role in gene regulation \cite{Kadauke2009}. Most of the asymptotic derivations are confirmed by Brownian simulations.\\
\textit{ \bf The RCL polymer model.} We start with a linear polymer in dimension $d (d=3)$ consisting of $N$ monomers with positions $R=[r_1,r_2,...,r_N]^T$, connected sequentially by harmonic springs \cite{Doi1986} and we added connectors between random non-nearest neighboring (NN) monomer pairs (Fig. \ref{fig:Figure_1}A). The potential energy of the RCL polymer is the sum of the spring potential of linear backbone and that of random connectors
\beq\label{eq:RCLPotential}
\phi(R)= \frac{\kappa}{2}\sum^N_{n=2}(r_n-r_{n-1})^2 +\frac{\kappa}{2}\sum_\mathcal{G} (r_m-r_n)^2,
 \eeq    	
where $\kappa=dk_BT/b^2$ is the spring constant, $b$ the standard-deviation of the connector between connected monomers, $k_B$ is the Boltzmann's constant and $T$ the temperature. The ensemble $\mathcal{G}$ is composed of $N_c$ randomly chosen indices $m, n$ among the non-NN monomers. We define the connectivity fraction $0\leq \xi\leq 1$, as the fraction of connector numbers $N_L=\frac{(N-1)(N-2)}{2}$,
\beq\label{eq:numRandomConnectors}
N_c(\xi) = \lfloor \xi N_L\rfloor.
\eeq
For each polymer realization, we choose $N_c$ pairs from the possible $N_L$ NN monomers. The dynamics of the resulting polymer monomers (vector $\R$) is driven by Brownian motion and the field of force due to the potential energy \ref{eq:RCLPotential}, leading to the stochastic description
\beq\label{eq:matrixFormOfRCLsystem}
\frac{d\R}{dt} = -\frac{d}{b^2}D\left(M+B(\xi)\right)\R +\sqrt{2D}\frac{d\Om}{dt},
\eeq
where $D=\frac{k_BT}{\zeta}$ is the diffusion constant, $\zeta$ is the friction coefficient, $\Om$ are independent white noise with mean 0 and variance 1, $M$ is the $N\times N$ Rouse matrix \cite{Doi1986}
{\small
\beq \label{eq:rouseMatrix}
  M_{m,n}= \begin{cases}
  	-\sum_{j\neq m} M_{m,j}, & m=n;\\
    	 -1 & |m-n|=1;\\
    	0, & \text{otherwise}.
  \end{cases}
\eeq
}
For a given $\xi$, the square symmetric matrix $B(\xi)$ with random connectivity is defined by
{\small
\beq
&&B_{mn}(\xi) =
\begin{cases}
-1, & |m-n|> 1, \text{and connected};\\
-\sum_{i\neq j}^N B_{mj}(\xi), & m=n;\\
0, & \text{ otherwise.}
\end{cases}\nonumber
\eeq
}
To derive the steady-state properties of an ensemble of RCL polymers, we adopt a mean-field model where we replace the matrix $B(\xi)$ in Eq. \ref{eq:matrixFormOfRCLsystem} by its average $\langle B(\xi)\rangle$ (averaging over all configurations of non NN connected monomer pairs). We thus construct $\langle B(\xi)\rangle$, using the probability density of the monomer connectivity. For a fixed number of connector $N_c$, the probability that monomer $m$ has $k\leq(N-2)$ non-NN connections is obtained by choosing $k$ position in row $m$ of the matrix $B(\xi)$ (excluding the super- and sub- and the diagonal), and the remaining $Nc-k$ connectors in any row or column $n\neq m$:
\beq\label{eq:conditionalProbabilitykConnectors}
 Pr_m(k) =
\begin{cases}
\frac{C^{N_c(\xi)-k}_{N_L-(N-3)}C^k_{N-3}}{C^{N_c(\xi)}_{N_L}}, & 1<m<N;\\
\frac{C^{N_c(\xi)-k}_{N_L-(N-2)}C^k_{N-2}}{C^{N_c(\xi)}_{N_L}}, & m=1,N,
 \end{cases}
\eeq
where the binomial coefficient is $C_i^j=\frac{i!}{(i-j)! j!}$. This probability is the hyper-geometric distribution for the number of connections for monomer $m$. The mean number of connectors for each monomer is therefore
 \beq\label{eq:expectedNumberOfConnectors}
 \beta_m(\xi)=
 \begin{cases}
    \frac{(N-3)N_c(\xi)}{N_L}\approx (N-3)\xi, & 1<m<N;\\
    \frac{(N-2)N_c(\xi)}{N_L}\approx (N-2)\xi, & m=1,N.
\end{cases}
\eeq
Using the mean values in \ref{eq:expectedNumberOfConnectors}, we obtain the expression for the matrix $\langle B(\xi) \rangle$, with entries
\beq\label{eq:averageB}
 \langle B_{mn}(\xi)\rangle = \begin{cases}
    -\xi,     & |m-n|>1;\\
 	\beta_m(\xi), & m=n;\\
 	0,        & \text{otherwise},\\
\end{cases}
\eeq
which can be decomposed as the sum
\beq\label{eq:averageBmatrixRepresentation}
\langle B(\xi) \rangle =\xi(N \Id-\M- \1_N),
\eeq
where $\Id$ is the $N\times N$ identity matrix, and $\1_N$ is a $N\times N$ matrix of ones. To study the mean properties of the RCL polymer, we study the stochastic process \ref{eq:matrixFormOfRCLsystem} using the average matrix $\langle B(\xi)\rangle$.
\\
\noindent\textit{\bf Eigenvalues of the RCL polymer.}
To study the steady-state properties of system \ref{eq:matrixFormOfRCLsystem}, we diagonalize the averaged connectivity matrix $M+\langle B(\xi)\rangle$. Using Rouse normal coordinates $U=[u_0, u_1,..u_{N-1}]$ \cite{Doi1986}, defined as
\beq\label{eq:normalCoordinates}
 U = \V R,
\eeq
where
\beq\label{eq:rouseEigenvectors}
\V=(\alpha_p^n)
= \begin{cases}
	\sqrt{\frac{1}{N}},& {p=0;}\\
	\sqrt{\frac{2}{N}}\cos\left((n-\frac{1}{2})\frac{p\pi}{N}\right), & {\text{otherwise}}
\end{cases}
\eeq
is the Rouse orthonormal basis \cite{Doi1986}, which diagonalizes $M$:
\beq
VMV^T=\Lambda=diag(\lambda_0,\lambda_1,...,\lambda_{N-1}),
\eeq
where
\beq\label{eq:rouseEigenvalues}
\lambda_p = 4\sin^2\left(\frac{p\pi}{2N}\right), \quad p=0,..,N-1,
\eeq	
are the eigenvalues of the Rouse matrix. We obtain from \ref{eq:matrixFormOfRCLsystem} the mean-field equations
\beq\label{eq:systemODENormalCoordinates}
\frac{d\U}{dt} =  -\frac{d}{b^2}D\left[\Lambda + V	\langle B(\xi) \rangle V^T\right]\U +\sqrt{2D}\frac{d\E}{dt},
\eeq
where $\E = V\omega$ are independent white noises with mean 0 and variance 1. From \ref{eq:averageBmatrixRepresentation}, the matrix $\langle B(\xi)\rangle$ commutes with $M$ and therefore is diagonalizable using the same orthonormal basis $V$:
\beq\label{eq:diagonalGammaValues}
V\langle B(\xi) \rangle V^T= diag(\gamma_0(\xi),...,\gamma_{N-1}(\xi)).
\eeq
Using \ref{eq:averageBmatrixRepresentation} and \ref{eq:diagonalGammaValues}, we obtain the eigenvalues
\beq
\gamma_p(\xi) =
\begin{cases}
	   0, & p=0;\\
\xi(N-\lambda_p), & 1\leq p \leq N-1.
\end{cases}
\eeq
Finally, the eigenvalues of system \ref{eq:systemODENormalCoordinates} are the sum of eigenvalues of the Rouse matrix $M$ and $\langle B(\xi)\rangle$:
\beq \label{eq:eigenValuesRCLPolymer}
\chi_p(\xi) = \gamma_p(\xi)+\lambda_p= 	N\xi+4(1-\xi)\sin^2\left(\frac{p\pi}{2N}\right).
\eeq
The system \ref{eq:systemODENormalCoordinates} is decoupled and consists of an ensemble of $N-$independent equations. For $\xi=0$, we recover the Rouse polymer \cite{Doi1986}, whereas for $\xi=1$, we obtain a fully connected polymer, for which all eigenvalues equal to $N$ except for the first vanishing one. Using \ref{eq:eigenValuesRCLPolymer}, the potential energy of the RCL polymer is written in the form
\beq\label{eq:diagonalizedRCLpotential}
\phi_{\xi}(U)= \frac{\kappa}{2}\sum_{p=1}^{N-1}\chi_p(\xi) u_p^2.
\eeq
The statistics of the RCL system (relation \ref{eq:matrixFormOfRCLsystem}), can be recovered from \ref{eq:systemODENormalCoordinates} in the diagonalized form (expression \ref{eq:diagonalizedRCLpotential}), by scaling $\xi$ with the ratio of mean number of random connectors to the mean of total number of connectors:
\beq\label{eq:xiTransformation}
\xi^* = \xi\frac{N_c}{N+N_c}.
\eeq
We plotted in Fig.\ref{fig:Figure_1}B the eigenvalues \ref{eq:eigenValuesRCLPolymer} for RCL polymers, for $N=50$ monomers, and $N_c$=5, 25 and 50 added random connectors.\\
\noindent\textit{\bf Encounter probability (EP) between monomers of the RCL polymer.}
The RCL polymer belongs to the class of generalized Gaussian chain models \cite{Sokolov2003,Gurtovenko2005, Jespersen2000}, for which the EP between any two monomers $m$ and $n$ at equilibrium is given by
\beq\label{eq:encounterProbInRandomLoopModel}
P_{m,n}(\xi)= \left(\frac{d}{2\pi \sigma_{m,n}^2(\xi)}\right)^{\frac{d}{2}}.
\eeq
To compute expression \ref{eq:encounterProbInRandomLoopModel}, we estimate the variance $\sigma_{m,n}^2(\xi)$ in normal coordinates (Eq. \ref{eq:normalCoordinates}):
\beq\label{eq:RCLvarianceSumInNormalCoordinate}
	\sigma_{m,n}^2(\xi)= \langle (r_m-r_n)^2\rangle=\sum_{p=0}^{N-1} \left(\alpha_p^m -\alpha_p^n\right)^2\langle u_p^2(\xi) \rangle.\nonumber
\eeq
From the Ornstein-Uhlenbeck equations \ref{eq:systemODENormalCoordinates} \cite{Schuss2009}, we obtain the time-dependent variance of the normal coordinates
\beq\label{eq:timeDependentNormVar}
\langle u_p^2(\xi)\rangle = \frac{b^2}{\chi_p(\xi)}\left(1-\exp\left(-\frac{2D\chi_p(\xi)t}{b^2}\right) \right).
\eeq
We define the hierarchy of relaxation times $\tau_0\geq \tau_1(\xi) \geq.. \tau_{N-1}(\xi)$,  with
 \beq\label{eq:relaxationTimes}
\tau_p(\xi) = \frac{b^2}{2D\chi_p(\xi)},
 \eeq
where the slowest time $\tau_0(\xi)$ corresponds to the diffusion of the center of mass. At steady-state,
\beq\label{eq:ststVariance}
\langle u_p^2(\xi)\rangle  = \frac{b^2}{2(1-\xi)\left(y(N,\xi)-\cos\left(\frac{p\pi}{N}\right)\right)},
\eeq
where
\beq\label{eq:ySubstitution}
y(N,\xi) = 1+\frac{N\xi}{2(1-\xi)}.
\eeq
Replacing \ref{eq:rouseEigenvectors} and \ref{eq:ststVariance} into \ref{eq:RCLvarianceSumInNormalCoordinate}, we get
{\small
\beq\label{eq:varianceSum}
\sigma_{m,n}^2(\xi) = \frac{b^2}{N(1-\xi)} \sum_{p=0}^{N-1}\frac{\left(\cos \left(\frac{p(m-\frac{1}{2})\pi}{N}\right)-\cos\left(\frac{p(n-\frac{1}{2})\pi}{N}\right)\right)^2}{y(N,\xi) -\cos(\frac{p\pi}{N})}.\nonumber
\eeq
}
For $N\gg1$, the sum \ref{eq:varianceSum} is computed in the complex plane using the contour of the unit disk parameterized by $z= e^{ix}$
{\small
\beq \label{eq:varianceComplexIntegral}	
\sigma_{m,n}^2(\xi) = \frac{-b^2}{4\pi i(1-\xi)}\oint_{|z|=1} \frac{(z-z^{m+n} )^2(z^m-z^n)^2dz}{(z-\zeta_0(N,\xi))(z-\zeta_1(N,\xi)) z^{2(m+n)+1}},\nonumber
\eeq
}
where
\beq\label{eq:rclVarianceDenominatorRoots}
\zeta_0(N,\xi) = y(N,\xi)+\sqrt{y^2(N,\xi)-1}, \nonumber \\
	\zeta_1(N,\xi) = y(N,\xi) -\sqrt{y^2(N,\xi)-1}.
\eeq
When $\zeta_0(N,0)=1$, we recover the variance $\sigma_{m,n}^2(0)=b^2|m-n|$ of the Rouse chain ($N_c=0$) \cite{Doi1986}. The integrand in  \ref{eq:varianceComplexIntegral} is symmetric in $m$ and $n$ and has a pole of order $2(m+n)+1$ at $z=0$ and simple poles at $z=\zeta_0(N,\xi), z= \zeta_1(N,\xi)$. Because $y(N,\xi)\geq1$, we have $\zeta_0(N,\xi)\geq 1$, which is outside of the unit  disk $|z|=1$, and $\zeta_1(N,\xi)\leq 1,$ for all N , $\xi\geq0$. The pole $\zeta_0(N,\xi)$ is not inside the disk and does not contribute in the calculation of the residues of \ref{eq:varianceComplexIntegral}. For $\xi>0$,  we obtain an exact expression for the variance
{
\small
\beq\label{eq:varianceExactForm}
\sigma_{m,n}^2(\xi) =
\begin{cases}
\frac{b^2}{(\zeta_0(N,\xi)-\zeta_1(N,\xi))(1-\xi)}\big(\frac{(\zeta_0^{m-n}(N,\xi) -1)^2-2\zeta_0^{m+n-1}(N,\xi)}{\zeta_0^{2m-1}(N,\xi)} \\ +2\big),  m \geq n;\\
	\frac{b^2}{(\zeta_0(N,\xi)-\zeta_1(N,\xi))(1-\xi)} \big(\frac{(\zeta_0^{n-m}(N,\xi) -1)^2-2\zeta_0^{m+n-1}(N,\xi) }{\zeta_0^{2n-1}}  \\  +2\big),  m<n.
\end{cases}
\eeq
}
For $\xi\ll 1$, the variance \ref{eq:varianceExactForm} is asymptotically given by
\beq\label{eq:varianceExpApproximation}
	\sigma_{m,n}^2(\xi) \approx \frac{b^2}{\sqrt{N\xi}}\left(1-\exp(-|m-n|\sqrt{N\xi}) \right).
\eeq
Using \ref{eq:varianceExactForm} and \ref{eq:encounterProbInRandomLoopModel}, we obtain a novel expression for the steady-state  encounter probability $P_{m,n}(\xi)$ between any two monomers. We compare the EP obtained from Brownian simulations of RCL polymer for $N=20,50$ with the analytical formula \ref{eq:encounterProbInRandomLoopModel} for $N_c=25$ connectors (Fig. \ref{fig:Figure_2}(a)), which shows a very good agreement.

\noindent\textit{ \bf Mean square radius of gyration (MSRG) of the RCL polymer.}
The MSRG $\langle R_g^2(\xi)\rangle$ characterizes the size of the RCL polymer and can be computed from the variance \ref{eq:varianceExactForm} as
\beq\label{eq:meanSquareRadiusRCLUsingTheVariance}
\displaystyle
\langle R_G^2(\xi) \rangle =  \frac{1}{N^2}\sum_{m=1}^N \sum_{n=1}^m \sigma_{m,n}^2(\xi).
\eeq
Using \ref{eq:varianceExactForm} and \ref{eq:meanSquareRadiusRCLUsingTheVariance}, with the notations $\zeta_0 = \zeta_0(N,\xi), \quad \zeta_1 = \zeta_1(N,\xi)$, we obtain
{\small
\beq\label{eq:MSRG}
     &&\langle R_G^2(\xi)\rangle =
     \frac{b^2}{N^2 (1-\xi)(\zeta_0-\zeta_1)}\Big[\frac{(1+2\zeta_0)N(1+N)}{2\zeta_0} \\
     &&+ \frac{N(2(1+\zeta_0)^2-\zeta_0^3)}{1-\zeta_0^2}-\frac{\zeta_0^3(1-\frac{1}
     {\zeta_0^{2N}})}{(1-\zeta_0^2)^2}+\frac{2(1+\zeta_0)
     (1-\frac{1}{\zeta_0^N})}{(1-\zeta_0)^2}\Big].\nonumber
\eeq
}
When $N_c\ll \frac{N^2}{2}$, we obtain the asymptotic expansion
\beq\label{eq:MSRGapproximation}
\langle R_G^2(\xi)\rangle \approx \frac{3b^2}{4}\sqrt{\frac{1}{N\xi}}.
\eeq

\begin{figure}[htbp!]
 \centering
 \includegraphics[width=1\linewidth]{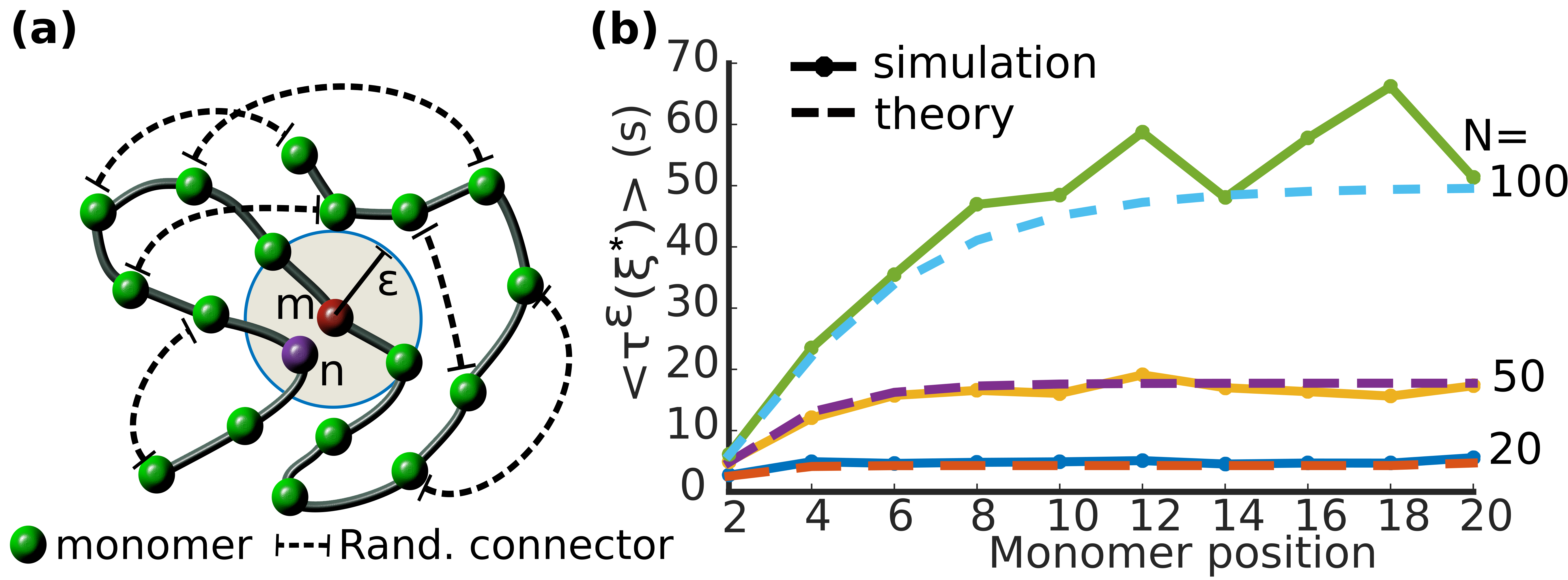}
 \caption{Transient properties of the RCL polymer \textbf{(a)} Two monomers $m$ (red) and $n$ (purple) of the RCL polymer meet when they enter a sphere of radius $\epsilon$. Random connectors (dashed arrows) are added to a linear Rouse backbone
 	\textbf{(b)} Stochastic simulations (marked by a dot) MFET between monomer 1 and monomers 2-20 of RCL polymers with $N=20$ (blue), 50 (yellow) and 100 (green) monomers, with $N_c=25$ random connectors, are in good agreement with the analytical formula (Eq. \ref{eq:MFETanalytical}, dashed). Parameters: $\epsilon=b/10, D=1, b=\sqrt{3}, \Delta t=0.01s$, the RCL system is \ref{eq:matrixFormOfRCLsystem} (we used Eq. \ref{eq:MFETanalytical} with $\xi^*$, Eq. \ref{eq:xiTransformation}).}
 	\label{fig:Figure_3}
  \end{figure}

In Fig. \ref{fig:Figure_1}(d), we compare $\langle R_G^2(\xi)\rangle$ computed from Brownian simulations of $N=20,50$, and 100 monomers, with $N_c\in [5,50]$ added random connectors, with the asymptotic formula \ref{eq:MSRG} and both agrees.\\
\textit{\bf Mean Square Displacement (MSD) of a single monomer of the RCL polymer.}
 Using the normal coordinates \ref{eq:normalCoordinates} in dimension $d$ the MSD of monomers in the RCL polymer is
 \beq\label{eq:msdSum}
 && \langle r_m^2(t) \rangle =
 \langle \left(\sum_{p=0}^{N-1}{\alpha_p^mu_p(t) }\right)^2\rangle = 2dD_{cm}t+ \\
 &&\frac{2db^2}{N}\sum_{p=1}^{N-1} {\frac{\cos^2\left(\frac{p\pi(m-\frac{1}{2})}{N}\right)\left(1-\exp(-\frac{2D\chi_p(\xi)t}{b^2}) \right)}{\chi_p(\xi)}},\nonumber
 \eeq
where $D_{cm}= \frac{D}{N}$. Averaging over all monomers and approximating the sum in \ref{eq:msdSum} by an integral for $N\gg1$, we obtain
\beq\label{eq:msdRCL}
  \langle \langle r_m^2(t) \rangle \rangle 
= 2dD_{cm}t +\frac{db^2 Erf[\sqrt{2dDN\xi t/b^2}]}{2\sqrt{N\xi(1-\xi)}},
 \eeq
where $Erf[t]$ is the error function. Equation \ref{eq:msdRCL} characterizes the MSD for intermediate time scale $\tau_{N-1}(\xi)\ll t\ll \tau_{1}(\xi)$. For short time scale $t\ll \tau_{N-1}(\xi)$, the MSD is given by
\beq
\langle \langle r_m^2(t)\rangle \rangle &&=\frac{db^2\int_0^{\sqrt{2dDN\xi t/b^2}}\exp(-g^2)dg}{\sqrt{\pi N\xi (1-\xi)}}\\
    &&\approx \frac{b \sqrt{2dDt}}{\sqrt{\pi (1-\xi)}}\left(1-\frac{\exp(-2dDN\xi t/b^2)}{2}\right).\nonumber
\eeq
For $N\xi\gg1$, the MSD behaves like
\beq
\langle \langle r_m^2(t)\rangle \rangle \propto \frac{db\sqrt{dDt}}{\sqrt{2\pi (1-\xi)}}.
\eeq
We conclude that the homogeneous behavior of MSD for the RCL polymer model gives an anomalous exponent $\alpha=0.5$, similar to the Rouse model. Finally, for long time scales ($t\gg\tau_1(\xi)$), (slow diffusion of the polymer's center of mass), the error function in \ref{eq:msdRCL} is almost constant and therefore
\beq
\langle \langle r_m(t)^2\rangle \rangle= 2dD_{cm}t +\frac{db^2}{2\sqrt{N\xi(1-\xi)}}.
\eeq
\noindent\textit{\bf Mean First Encounter Time (MFET) $\langle \tau^\epsilon (\xi) \rangle$ between monomers of the RCL polymer.}
We compute here the mean time for two monomers of the RCL polymer to enter for the first time in a ball of radius $\epsilon>0$, at which they can possibly interact to form a chemical bond (Fig. \ref{fig:Figure_3}(a)). The MFET for both the Rouse and beta \cite{Amitai2013} polymer were computed (see \cite{Amitai2012}) from the first eigenvalue $\lambda_0^\epsilon$ of the Fokker-Planck operator associated to the stochastic equation \ref{eq:systemODENormalCoordinates}, so that
\beq\label{eq:MFETsingleTADDefinition}
\langle \tau^{\epsilon}(\xi)\rangle\approx \frac{1}{D\lambda_0^{\epsilon}(\xi)}.
\eeq
The first order approximation in $\epsilon$ is given by \cite{Amitai2012}
\beq\label{eq:firstEigenValueFKOperator}
\lambda_0^{\epsilon}(\xi) = \frac{4\pi \epsilon\int_{C-P}e^{-\phi_\xi(U)}dU}{|\tilde{\Omega}(\xi)|}+O(\epsilon^2),
\eeq
where $\phi_\xi(U)$ is the diagonalized potential \ref{eq:diagonalizedRCLpotential}, $|\tilde{\Omega}(\xi)|$ is the integral over the entire RCL configuration space
\beq\label{eq:configurationSPaceRCL}
|\tilde{\Omega}(\xi)|=\int e^{-\phi_\xi(U)} dU =\left( \frac{(2\pi)^{N-1}}{\prod_{p=1}^{N-1}   \kappa \chi_p(\xi)}\right) ^{\frac{d}{2}}.
\eeq
The integral over $C-P$ in \ref{eq:firstEigenValueFKOperator} is computed over the space of closed RCL polymer ensemble, with fixed connector between monomers $m$ and $n$ and additional $N_c(\xi)$ random connectors. A direct computation gives
{\small
\beq \label{eq:closedPolymerEnsemble}
\int_{C-P}e^{-\phi_\xi(U)}dU=
(2\pi)^{\frac{(N-2)d}{2}}\left(\frac{\kappa b^2\prod_{p=1}^{N-1}(\kappa \chi_p(\xi))^{-1}}{
\sigma_{m,n}^2(\xi)}\right)^{\frac{d}{2}}.
\eeq
}
Using relations \ref{eq:configurationSPaceRCL} and \ref{eq:closedPolymerEnsemble} in \ref{eq:MFETsingleTADDefinition}, we obtain the MFET between any two monomers $m$ and $n$ of the RCL polymer for a given connectivity fraction $\xi$ in dimension $d=3$:
\beq \label{eq:MFETanalytical}
\langle \tau^{\epsilon}_{m,n}(\xi)\rangle
=\frac{1}{4\pi D\epsilon }\left(\frac{2\pi\sigma_{m,n}^2(\xi)}{\kappa b^2}\right)^{\frac{3}{2}},
\eeq
Using \ref{eq:varianceExpApproximation} into \ref{eq:MFETanalytical}, we obtain the approximation
\beq\label{eq:mfetApproximation}
\langle \tau^{\epsilon}_{m,n}(\xi)\rangle \approx \frac{b^2\left(1-\exp(-|m-n|\sqrt{N\xi}) \right)^{d/2}}{4\sqrt{N\xi}\pi D\epsilon(\kappa b^2)^{d/2}}  +\mathcal{O}(N\xi),\nonumber
\eeq
where $|m-n|\ll N$, and $\xi\ll 1$. The analytical formula \ref{eq:MFETanalytical} agrees with Brownian simulations of the MFET for RCL polymer (Eq. \ref{eq:matrixFormOfRCLsystem}) with $N=20,50,$ and 100 monomers, and $N_c=25$ added random connectors (Fig. \ref{fig:Figure_3}(b)).

\noindent\textit{\bf Applications of the RCL polymer model.}
We derived here several analytical formula for the steady-state variance, encounter probability, the radius of gyration, mean-square displacement and the mean first encounter time of the RCL polymer model. These formula can be used to extract parameters from chromatin conformation in CC experiments \cite{Dixon2012,Nora2012}. In particular, using formula \ref{eq:encounterProbInRandomLoopModel}, it is possible to fit the empirical encounter probability obtained from experimental data to extract the connectivity fraction $\xi$. This parameter has a direct interpretation and represents the mean number of cross-links, that can be mediated by CTCF molecules present in a genomic region. The parameter $\xi$ depend on the coarse-grained scale (see \cite{Shukron2017}). The extracted parameter $\xi$ can then be used to estimate the radius of gyration (Eq. \ref{eq:MSRG}) of any region of interest. This radius characterizes the size of the genomic region, at least relative to other genomic segments, hence providing insightful information about the local organization of the chromatin in the cell nucleus.

To demonstrate out methodology, we coarse-grained the 5C data reported in \cite{Nora2012} of male neuronal progenitors NPC-E14 cells, replicate 1, TAD H, containing 679 kbp, at a scale of 3kbp, resulting in $N=226$ monomers. We fit the EP (Eq.\ref{eq:encounterProbInRandomLoopModel}) to 5C data of each of the 226 monomers and obtain the average connectivity $\xi=0.0022$, corresponding $N_c=56$ added connectors. We use the persistence length of $b=0.05$ $\mu m$. Substituting $N=226,b=0.05,\xi=0.0022$ in Eq. \ref{eq:MSRG}, we compute the radius of gyration to be 43 nm for TAD H. Thus, the 679 kbp TAD H is compacted in a sphere of volume $3.4\times 10^5$ $nm^3$ (2 bp per $nm^3$).

The structural information extracted from the static CC maps using the RCL polymer model was recently used to interpret the dynamics of single particle trajectories (SPT) \cite{Gasser2016,Bystricky2015,Amitai2017}. By fitting the MSD (Eq. \ref{eq:msdRCL}) to SPT data and by extracting the degree of connectivity $\xi$, we interpret the mean deviation of the loci dynamic from pure diffusion as the confined dynamics of the loci in a cross-links genomic environment \cite{Shukron2017,Weber2010PRE,Weber2010PRL}. We provided a direct formula to extract $\xi$, so that the simulations of \cite{Amitai2017} can now be bypassed. Finally, once the connectivity fraction is extracted, the mean first encounter time between any two monomers can be computed using formula \ref{eq:MFETanalytical}. Encounter times are key for understanding processes, such as mammalian X chromosome inactivation \cite{Nora2012} or non-homologous-end joining after DNA double-strand break \cite{Amitai2017,Amitai2012}.

	
%

	\end{document}